\documentclass[aps,prl,twocolumn,groupedaddress]{revtex4}
\usepackage[dvips]{graphicx,color} % Graphics and color
\usepackage{array,hhline,dcolumn} % Better table handling
\usepackage{rotating} % Rotate and sideways environments

\newcommand{\numu}{\nu_{\mu}}
\newcommand{\numubar}{\bar{\nu}_{\mu}}
\newcommand{\nue}{\nu_e}

\newcommand{\Elo}{E_\mathrm{lo}}
\newcommand{\Ehi}{E_\mathrm{hi}}

\begin{document}

% this adds line numbers to all text in the document (useful for editing)
%\pagewiselinenumbers

% APS preprint designation
\preprint{MiniBooNE-qe}

\title{Measurement of Muon Neutrino Quasi-Elastic Scattering on Carbon
% put some version info here
}
\date{\today}

\author{A.~A. Aguilar-Arevalo$^{5}$, A.~O.~Bazarko$^{12}$, S.~J.~Brice$^{7}$, 
        B.~C.~Brown$^{7}$, L.~Bugel$^{5}$, J.~Cao$^{11}$, L.~Coney$^{5}$,
        J.~M.~Conrad$^{5}$, D.~C.~Cox$^{8}$, A.~Curioni$^{16}$, 
        Z.~Djurcic$^{5}$, D.~A.~Finley$^{7}$, B.~T.~Fleming$^{16}$,       
        R.~Ford$^{7}$, F.~G.~Garcia$^{7}$, G.~T.~Garvey$^{9}$, C.~Green$^{7,9}$,
        J.~A.~Green$^{8,9}$, T.~L.~Hart$^{4}$, E.~Hawker$^{15}$,
        R.~Imlay$^{10}$, R.~A. ~Johnson$^{3}$, P.~Kasper$^{7}$, T.~Katori$^{8}$,
        T.~Kobilarcik$^{7}$, I.~Kourbanis$^{7}$, S.~Koutsoliotas$^{2}$, 
        E.~M.~Laird$^{12}$, J.~M.~Link$^{14}$, Y.~Liu$^{11}$, Y.~Liu$^{1}$, 
        W.~C.~Louis$^{9}$, K.~B.~M.~Mahn$^{5}$, W.~Marsh$^{7}$, 
        P.~S.~Martin$^{7}$, G.~McGregor$^{9}$, W.~Metcalf$^{10}$, 
        P.~D.~Meyers$^{12}$, F.~Mills$^{7}$, G.~B.~Mills$^{9}$,
        J.~Monroe$^{5}$, C.~D.~Moore$^{7}$, R.~H.~Nelson$^{4}$,
        P.~Nienaber$^{13}$, S.~Ouedraogo$^{10}$, R.~B.~Patterson$^{12}$,
        D.~Perevalov$^{1}$, C.~C.~Polly$^{8}$, E.~Prebys$^{7}$,
        J.~L.~Raaf$^{3}$, H.~Ray$^{9}$, B.~P.~Roe$^{11}$, 
        A.~D.~Russell$^{7}$, V.~Sandberg$^{9}$, R.~Schirato$^{9}$,
        D.~Schmitz$^{5}$, M.~H.~Shaevitz$^{5}$, F.~C.~Shoemaker$^{12}$,
        D.~Smith$^{6}$, M.~Sorel$^{5}$, P.~Spentzouris$^{7}$, I.~Stancu$^{1}$,
        R.~J.~Stefanski$^{7}$, M.~Sung$^{10}$, H.~A.~Tanaka$^{12}$,       
        R.~Tayloe$^{8}$, M.~Tzanov$^{4}$, R.~Van~de~Water$^{9}$, 
        M.~O.~Wascko$^{10}$, D.~H.~White$^{9}$, M.~J.~Wilking$^{4}$, 
        H.~J.~Yang$^{11}$, G.~P.~Zeller$^{5}$, E.~D.~Zimmerman$^{4}$ 
\\\smallskip
(The MiniBooNE Collaboration) 
\smallskip}\smallskip\smallskip

\affiliation{$^1$University of Alabama; Tuscaloosa, AL 35487 \\
               $^2$Bucknell University; Lewisburg, PA 17837 \\
             $^3$University of Cincinnati; Cincinnati, OH 45221\\
             $^4$University of Colorado; Boulder, CO 80309 \\
             $^5$Columbia University; New York, NY 10027 \\
             $^6$Embry-Riddle Aeronautical University; Prescott, AZ 86301 \\
             $^7$Fermi National Accelerator Laboratory; Batavia, IL 60510 \\
             $^8$Indiana University; Bloomington, IN 47405 \\
             $^9$Los Alamos National Laboratory; Los Alamos, NM 87545 \\
             $^{10}$Louisiana State University; Baton Rouge, LA 70803 \\
             $^{11}$University of Michigan; Ann Arbor, MI 48109 \\
             $^{12}$Princeton University; Princeton, NJ 08544 \\
             $^{13}$Saint Mary's University of Minnesota; Winona, MN 55987 \\
             $^{14}$Virginia Polytechnic Institute \& State University; Blacksburg, VA 24061\\
             $^{15}$Western Illinois University; Macomb, IL 61455 \\
             $^{16}$Yale University; New Haven, CT 06520\\
}

\begin{abstract}
The observation of neutrino oscillations is clear evidence for physics beyond 
the standard model. To make precise measurements of this phenomenon, neutrino 
oscillation experiments, including MiniBooNE, require 
an accurate description of neutrino charged current quasi-elastic (CCQE) 
cross sections to predict signal samples. Using a high-statistics sample of
$\numu$ CCQE events, MiniBooNE finds that a simple Fermi gas model, 
with appropriate adjustments, accurately characterizes the CCQE events 
observed in a carbon-based detector. The extracted parameters include an 
effective axial mass, $M_A^{\rm eff}=1.23\pm 0.20$ GeV, that describes the 
four-momentum dependence of the axial-vector form factor of the nucleon; and 
a Pauli-suppression parameter, $\kappa = 1.019 \pm 0.011$. Such a modified 
Fermi gas model may also be used by future accelerator-based experiments 
measuring neutrino oscillations on nuclear targets.
\end{abstract}

% Use showkeys class option if keyword
% display desired
\maketitle
%-----------------------------------------------------------------------------
% --- introduction

The recent observation of neutrino oscillations is strong evidence for
massive neutrinos and, therefore, for physics beyond the standard model.
Accelerator-based experiments searching for neutrino oscillations, such as
MiniBooNE~\cite{MB-PRL} and K2K~\cite{k2k-osc}, use charged current 
quasi-elastic (CCQE) interactions to search for the appearance of electron 
neutrinos ($\nue n \rightarrow e^- p$) in beams of muon neutrinos. The 
muon neutrino CCQE interaction ($\numu n \rightarrow \mu^- p$) thus provides 
a calibration for the neutrino beam and for the interaction cross section.  
In addition, such events dominate at energies between 200-2000~MeV where the 
oscillation searches are conducted. To ensure high event yields, these 
experiments use nuclear media (carbon or water) as the neutrino target; 
therefore, it is crucial to employ an accurate model of the CCQE interaction 
on nuclei.  In this Letter, we describe the model improvements developed
for the recent oscillation search from the MiniBooNE experiment~\cite{MB-PRL}.
The modified model describes this reaction remarkably well and should
be relevant for future accelerator-based neutrino oscillation searches.

%Charged current quasi-elastic (CCQE, $\numu \, n \rightarrow \mu^- \, p$) 
%events dominate neutrino interactions at energies between $200-2000$ MeV.
%Because the process has a large cross section and identifies both the incident 
%neutrino flavor and energy, it forms an ideal signal sample in neutrino 
%oscillation experiments. To ensure high event yields, typical experiments
%use nuclear media such as carbon, water, or iron for their 
%neutrino targets.  As a result, 
%a detailed understanding of quasi-elastic scattering on nuclear targets is 
%required. 

To model the scattering from nucleons confined in nuclei,
most neutrino oscillation experiments employ an event generator based on 
the relativistic Fermi gas (RFG) model~\cite{smith-moniz}.
Such models assume a flat nucleon momentum distribution up to some
Fermi momentum ($p_F$), assign a single value for the nucleon binding energy 
($E_B$) to account for the initial and final state total energies, 
%assign a nuclear binding energy ($E_B$) to account
%for nuclear interactions in the initial and final states, 
and utilize standard nucleon vector and axial-vector on-shell form factors.
Many of these model parameters may be inferred from existing data; for 
example, $p_F$, $E_B$, and the vector form factors can be determined from 
elastic electron scattering data~\cite{eb-pf,bba}.  Despite providing these 
constraints, electron data yield limited information on the axial-vector 
form factor of the nucleon and the CCQE cross section at very low four-momentum
transfer ($Q^2$). Present knowledge of the axial-vector form factor has been 
informed largely by past neutrino experiments, but these suffer from low 
statistics and were performed using predominantly deuterium 
targets~\cite{past-qe}. Since these early measurements, neutrino experiments 
have encountered difficulties describing their data at low $Q^2$, where 
nuclear effects are largest, and have often measured axial-vector form factor 
parameters above some minimum $Q^2$ value. 

The MiniBooNE experiment has collected the largest sample of low
energy muon neutrino CCQE events to date. We describe here the use of such 
events in tuning the RFG model to better describe quasi-elastic scattering on 
nuclear targets. The analysis fits the reconstructed $Q^2$ distribution
of the MiniBooNE CCQE data in the region $0<Q^2<1$~GeV$^2$ to a
simple RFG model~\cite{smith-moniz} with two adjustable parameters:
the axial mass, $M_A$, appearing in the axial-vector form factor; and 
$\kappa$, a parameter that adjusts the level of Pauli-blocking at low values 
of $Q^2$. The best-fit model results in a good description of the data across 
the full kinematic phase space including the low-$Q^2$ region. This technique 
is crucial to the MiniBooNE oscillation search~\cite{MB-PRL} as it is used to
predict the $\nue$ CCQE oscillation events based on the constraints provided 
by the high-statistics MiniBooNE $\numu$ CCQE sample.

%---- experiment

The Fermilab Booster neutrino beam, optimized for the MiniBooNE 
oscillation search, is particularly suited for 
investigation of low energy neutrino interactions. The Fermilab Booster
provides 8.89 GeV/$c$ protons which collide 
with a 71 cm long beryllium target inside a magnetic horn. The horn focuses 
positively charged pions and kaons produced in these collisions, which can 
subsequently decay in a 50~m long decay region, yielding an intense flux 
of muon neutrinos. A \textsc{geant4}-based~\cite{GEANT4} beam simulation uses 
a parametrization~\cite{flux-param} of pion production cross sections 
based on recent measurements from the HARP~\cite{harp} and E910~\cite{e910}
experiments, along with a detailed model of the beamline geometry to predict 
the neutrino flux as a function of neutrino energy and flavor. The resulting 
flux of neutrinos at the MiniBooNE detector is predicted to be $93.8\%$ 
($5.7\%$) $\numu$ ($\numubar$) with a mean energy of $\sim700$~MeV. 
Because $99\%$ of the flux lies below 2.5~GeV, the background from high 
multiplicity neutrino interactions is small. 
Approximately $40\%$ of the total events at MiniBooNE are predicted to 
be $\numu$ CCQE, of which $96\%$ result from pion decays in the beam.

The MiniBooNE detector is a spherical tank of inner radius 610 cm filled with 
800 tons of mineral oil (CH$_2$), situated 541 meters downstream of the 
proton target. An optical barrier divides the detector into two regions,
an inner volume with a radius of 575 cm and an outer volume 35 cm thick.
The inner region of the tank houses 1280 inward-facing 
8 inch photomultiplier tubes (PMTs), providing $10\%$ photocathode coverage. 
The outer region is lined with 240 pair-mounted PMTs which provide a veto 
for charged particles entering or leaving the tank. 
Muons produced in CCQE interactions emit primarily Cherenkov light with 
a small amount of scintillation light. A large number of muons stop 
and decay in the main detector volume. The muon kinetic energy resolution 
is $7\%$ at 300 MeV and the angular resolution is $5^{\circ}$. The response 
of the detector to muons is calibrated using a dedicated muon tagging system 
that independently measures the muon energy for cosmic ray muons ranging 
up to 800~MeV.

Neutrino interactions within the detector are simulated with the v3 
\textsc{nuance}
event generator~\cite{nuance}.  This program provides the framework for  
tuning the CCQE cross section parameters (described below) and predicts
backgrounds to the sample, including neutrino induced
single pion production events (CC $1\pi$). Pion interactions in the 
nucleus and photon emission from nuclear de-excitation in \textsc{nuance} are 
tuned to reproduce MiniBooNE and other~\cite{ashery} data. 
A \textsc{geant3}-based~\cite{GEANT3} detector model (with 
\textsc{gcalor}~\cite{gcalor} 
hadronic interactions) simulates the detector response to particles produced 
in neutrino interactions. The simulation of light production and propagation
in mineral oil has been tuned using external small-sample 
measurements~\cite{ieee}, muon decay electrons (also used to calibrate the 
energy scale), and recoil nucleons from neutrino neutral current (NC) elastic 
scattering events. The predicted events are additionally overlaid with events 
measured in a beam-off gate, in order to incorporate backgrounds from natural
radioactivity and cosmic rays into the simulated data.

%----- selection

Because of the low energy neutrino beam and MiniBooNE detector capabilities, 
the identification of $\numu$ CCQE interactions relies solely on 
the detection of the primary muon and associated decay electron in these 
events:

\vspace{-0.1in}
\[
  \nu_\mu+n \rightarrow \mu^{-}+p, \hspace{0.2in}
  \mu^{-} \rightarrow e^{-} + \nu_\mu + \bar{\nu_e}.
\]
This simple selection is highly effective for several reasons.
First, the efficiency for detecting the decay of the $\mu^-$ produced in
such events is high, 83\%. The losses are due to muon capture on
carbon ($8\%$~\cite{muoncap}) and insufficient decay time or energy of the 
decay electron ($10\%$). Second, the CC $1\pi^+$ contamination 
is significantly reduced by requiring a single decay electron, since CC 
$1\pi^+$ events typically yield two decay electrons, one each from the 
primary muon and the $\pi^+$ decay chains. The exceptions are cases in which 
the primary $\mu^-$ is captured or, more likely, the $\pi^+$ is either 
absorbed or undergoes a charge-changing interaction in the target nucleus 
or detector medium. Each of these processes is included in the detector 
simulation. Finally, by avoiding requirements on the outgoing proton 
kinematics, the selection is inherently less dependent on nuclear models.

Timing information from the PMTs allows the light produced by the initial
neutrino interaction (first ``sub-event'') to be separated from light 
produced by the decay electron (second sub-event).  
The time and charge response of the PMTs is used to reconstruct the position, 
kinetic energy, and direction vector of the primary particle within each 
sub-event. Once separated into sub-events, we require that the first sub-event
(the neutrino interaction) must occur in coincidence with a beam pulse, have a 
reconstructed position $<500$~cm from 
the center of the detector, possess $<6$ veto-PMT hits to ensure
containment, and have $>200$ main-PMT hits to avoid electrons from cosmic ray
muon decays. The second sub-event (the $\mu^-$ decay electron) must have 
$<6$ veto-PMT hits and $<200$ main-PMT hits.  Subsequent cuts specifically 
select $\numu$ CCQE events and discriminate against CC $1\pi^+$ backgrounds. 
First, events must contain exactly two sub-events. Second, the distance 
between the electron vertex and muon track endpoint must be less than 100~cm, 
ensuring that the decay electron is associated with the muon track.

A total of 193,709 events pass the MiniBooNE $\numu$ CCQE selection 
criteria from $5.58\times10^{20}$ protons on target collected between 
August 2002 and December 2005. The cuts are estimated to be $35\%$ efficient 
at selecting $\numu$ CCQE events in a 500 cm radius, with a CCQE purity 
of $74\%$. The $35\%$ efficiency is the product of a $50\%$ probability
for containing events within the tank, the aforementioned $83\%$ muon decay
detection efficiency, and an $85\%$ efficiency for the electron vertex to
muon endpoint requirement.

%The predicted  backgrounds are: $74.8\%$ CC $1\pi^+$, $15.0\%$ 
%CC $1\pi^0$, $4.0\%$ NC $1\pi^{\pm}$, $2.6\%$ CC multi-$\pi$, $0.9\%$ 
%NC elastic, $0.8\%$ $\numubar$ CC $1\pi^-$, $0.8\%$ NC $1\pi^0$, $0.6\%$ 
%$\eta$/$\rho$/$K$ production,  and $0.5\%$ deep inelastic scattering and 
%other events~\cite{nuance}. Because pions can be absorbed via final state 
The predicted  backgrounds are: $75\%$ CC $1\pi^+$, $15\%$ 
CC $1\pi^0$, $4\%$ NC $1\pi^{\pm}$, $3\%$ CC multi-$\pi$, $1\%$ 
NC elastic, $1\%$ $\numubar$ CC $1\pi^-$, $1\%$ NC $1\pi^0$, $<1\%$ 
$\eta$/$\rho$/$K$ production,  and $<1\%$ deep inelastic scattering (DIS)
and other events~\cite{nuance}. In the analysis, cross section uncertainties
of $25\%$, $40\%$, and $25\%$ are assumed on the 1$\pi$, multi-$\pi$
plus $\eta$/$\rho$/$K$ production, and DIS backgrounds, respectively. 
Because pions can be absorbed via final state 
interactions in the target nucleus, a large fraction of the background 
events look like CCQE events in the MiniBooNE detector. ``CCQE-like'' events, 
all events with a muon and no pions in the final state, are predicted to be 
$84\%$ of the sample after cuts.

%------ results

\begin{figure}
\includegraphics[bb=16 8 564 405,height=2.6in]{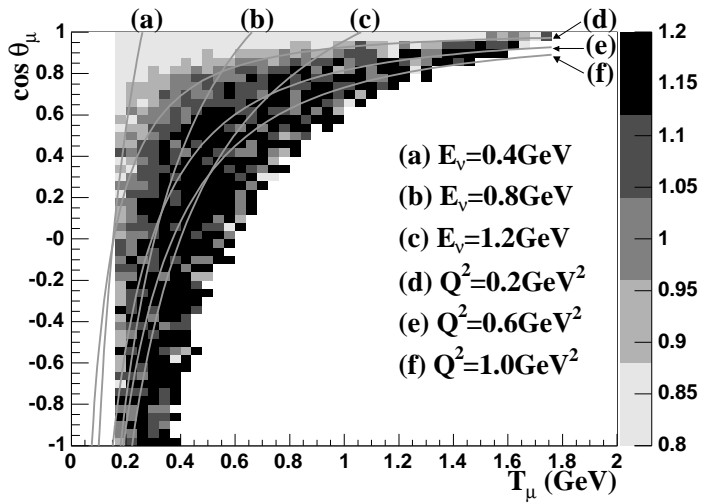}
\caption{Ratio of MiniBooNE $\numu$ CCQE data/simulation as a function of 
reconstructed muon angle and kinetic energy. The prediction 
is prior to any CCQE model adjustments and has been normalized to 
the data. The $\chi^2/\mathrm{dof}=79.5/53$. The ratio forms a 2D surface 
whose values are represented by the gray scale, shown on the right. If the 
simulation modeled the data perfectly, the ratio would be unity everywhere. 
Contours of constant $E_\nu$ and $Q^2$ are overlaid, and only bins with $>20$ 
events in the data are plotted.}
\label{fig:prl_fig1}
\end{figure}

The observables in the MiniBooNE $\numu$ CCQE sample are the muon kinetic 
energy $T_\mu$, and the muon angle with respect to the neutrino beam direction
$\theta_\mu$. The high-statistics MiniBooNE data sample allows us to 
verify the simulation in two dimensions. Figure~\ref{fig:prl_fig1} shows the 
level of agreement between the shape of the data and simulation in the CCQE 
kinematic quantities before any CCQE cross section model adjustments. 
For this comparison, the simulation assumes the RFG 
model as implemented in \textsc{nuance}~\cite{smith-moniz,nuance}, with  
$E_B=34$~MeV~\cite{eb-pf}, $p_F=220$~MeV/$c$~\cite{eb-pf}, updated
non-dipole vector form factors~\cite{bba}, and a non-zero pseudoscalar
form factor~\cite{pseudo}. The axial-vector form factor 
is assumed to have a dipole form as a function of $Q^2$ 
with one adjustable parameter, $M_A$, the so-called ``axial mass'',
$F_A(Q^2) = g_A/(1+Q^2/M_A^2)^2$.
%
%\vspace{-0.2in}
%\begin{equation}
%   F_A(Q^2) = g_A/(1+Q^2/M_A^2)^2.
%\end{equation}

%\noindent
The simulation shown in Fig.~\ref{fig:prl_fig1} specifically assumes 
$g_A=1.2671$~\cite{ga} and $M_A=1.03$ GeV~\cite{past-ma}. These model 
parameters are common defaults in most neutrino simulations. The figure 
shows that the disagreement between data and simulation  follows lines of 
constant $Q^2$ and not $E_\nu$.  This supports the assumption that the 
data/model disagreement is not due to a mis-modeling of the incoming 
neutrino energy spectrum but an inaccuracy in the simulation of the CCQE 
process itself. We also explicitly assume no $\numu$ disappearance due to 
oscillations.

Guided by indications that the data-model discrepancy is only a function 
of $Q^2$, we have modified the existing $\numu$ CCQE model 
rather than introduce more drastic changes to the cross section calculation. 
This approach works well and requires adjustment of only two 
parameters: $M_A$ and $\Elo$. The parameter $\Elo$ effectively controls the 
effect of Pauli-blocking. It is the lower bound of integration over initial 
state nucleon energy and appears within the RFG model together with an upper 
bound $\Ehi$: 
\begin{eqnarray}
       \Ehi=\sqrt{p_F^2+M_n^2}, \hspace{0.05in}  %  \label{eqn:ehi} \\ 
       \Elo=\sqrt{p_F^2+M_p^2} - \omega+E_B,     %  \label{eqn:elo}
\end{eqnarray} 
where $M_n$ is the target neutron mass, $M_p$ is the outgoing proton mass, 
and  $\omega$ is the energy transfer.
In the RFG model, $\Ehi$ is the energy of an initial nucleon on the Fermi 
surface and $\Elo$ is the lowest energy of an initial nucleon that leads to a 
final nucleon just above the Fermi momentum (and thus obeying the exclusion 
principle in the final state).  In practice, a simple scaling of $\Elo$ was 
implemented in the MiniBooNE CCQE data fit via 
$\Elo=\kappa(\sqrt{p_F^2+M_p^2} - \omega+E_B)$. The parameter $\kappa$ adds a 
degree of freedom to the RFG model which can describe the smaller
cross section observed in the data at low momentum transfer and 
is likely compensating for the naive treatment of Pauli-blocking in the RFG 
model. 
%Adjustment of both parameters, $M_A$ and $\kappa$, is likely compensating 
%for an inadequate nucleon momentum distribution in the RFG model.

The adjusted RFG model is then fit to the shape of the reconstructed $Q^2$ 
distribution in the MiniBooNE $\numu$ CCQE data:

\vspace{-0.2in}
\begin{eqnarray}
    Q^2 = -q^2 = -m_\mu^2 + 2 E_\nu(E_\mu - p_\mu \cos\theta_\mu) > 0,
\end{eqnarray}

\noindent
where $m_\mu$ is the muon mass, $E_\mu$ ($p_\mu$) is the reconstructed muon 
energy (momentum), and $\theta_\mu$ is the reconstructed muon scattering angle. 
The reconstructed neutrino energy $E_\nu$ is formed assuming the target 
nucleon is at rest inside the nucleus:

\vspace{-0.1in}
\begin{eqnarray}
     E_\nu = \frac{2(M_n - E_B)E_\mu - (E_B^2 - 2M_n E_B + m_\mu^2 +
             \Delta M^2)}
             {2\:[(M_n - E_B) - E_\mu + p_\mu \cos\theta_\mu]},
\end{eqnarray}

\noindent
where $\Delta M^2 = M_n^2 - M_p^2$ 
and $E_B>0$. A small correction is applied to $E_\nu$ in both 
data and simulation to account for the biasing effects of Fermi-smearing.
This procedure, while yielding a more accurate $E_\nu$ estimate, has a
negligible impact on the $Q^2$ fit to MiniBooNE CCQE data. These expressions, 
with reconstructed muon kinematics, yield an $E_\nu$ resolution of $11\%$ 
and a $Q^2$ resolution of $21\%$ for CCQE events.

\begin{figure}
\includegraphics[bb=18 8 564 405,height=2.6in]{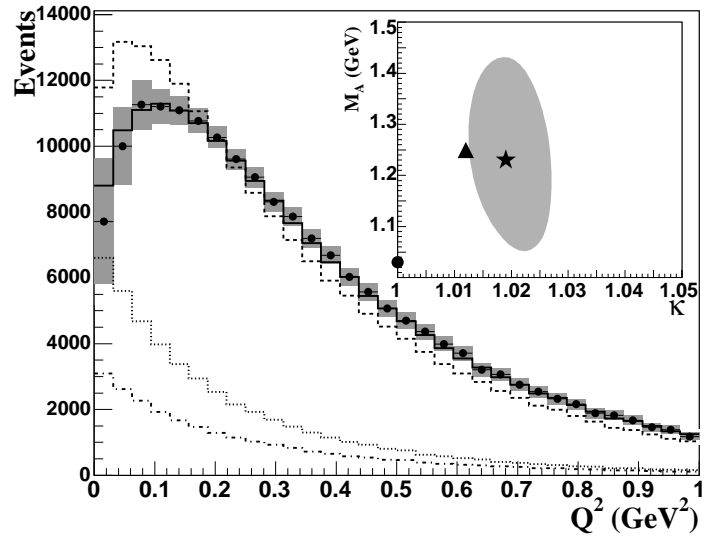}
\caption{Reconstructed $Q^2$ for $\numu$ CCQE events including
systematic errors. The simulation, before (dashed) and after (solid) the fit, 
is normalized to data. The dotted (dot-dash) curve shows backgrounds that 
are not CCQE (not ``CCQE-like''). The inset shows the 1$\sigma$ CL contour 
for the best-fit parameters (star), along with the starting values (circle), 
and fit results after varying the background shape (triangle).}
\label{fig:prl_fig2}
\end{figure}

The model parameters $M_A$ and $\kappa$ are obtained from a least-squares 
fit to the measured data in 32 bins of reconstructed $Q^2$ from 0 to 1~GeV$^2$.
All other parameters of the model are held fixed to the values
listed previously, and a complete set of correlations between 
systematic uncertainties is considered. The total prediction is normalized 
to the data for each set of parameter values. Thus, the procedure is sensitive
only to the shape of the $Q^2$ distribution, and any changes in the total 
cross section due to parameter variation do not impact the quality of fit. 
The $Q^2$ distributions 
of data and simulation before and after the fitting procedure are shown in 
Figure~\ref{fig:prl_fig2}. The $\chi^2/\mathrm{dof}$ of the fit is 32.8/30 
and the parameters extracted from the MiniBooNE $\numu$ CCQE data are: 
\begin{eqnarray}
  M_A^{\rm eff} &=& 1.23 \pm 0.20 \hspace{0.1in} \mathrm{GeV}; \\
  \kappa &=& 1.019 \pm 0.011.
\end{eqnarray}

\noindent
While normalization is not explicitly used in the fit, the new model 
parameters increase the predicted rate of $\numu$ CCQE events at MiniBooNE 
by $5.6\%$. The ratio of detected events to predicted, with the best-fit
CCQE model parameters, is $1.21 \pm 0.24$.
%The observed rate of $\numu$ CCQE events in the MiniBooNE 
%data is $1.21 \pm 0.24$ when compared to the prediction with these best-fit 
%CCQE model parameters.

In general, varying $M_A$ allows us to reproduce the high $Q^2$ behavior of
the observed data events. A fit for $M_A$ above $Q^2>0.25$ GeV$^2$ yields 
consistent results, $M_A^{\rm eff} = 1.25 \pm 0.12$ GeV. However, fits varying 
only $M_A$ across the entire $Q^2$ range leave considerable disagreement at 
low $Q^2$ ($\chi^2/\mathrm{dof}=48.8/31$). The Pauli-blocking parameter 
$\kappa$ is instrumental here, enabling this model to match the behavior 
of the data down to $Q^2=0$ (Figure~\ref{fig:prl_fig2}). 

\begin{figure}
\includegraphics[bb=16 8 564 405,height=2.6in]{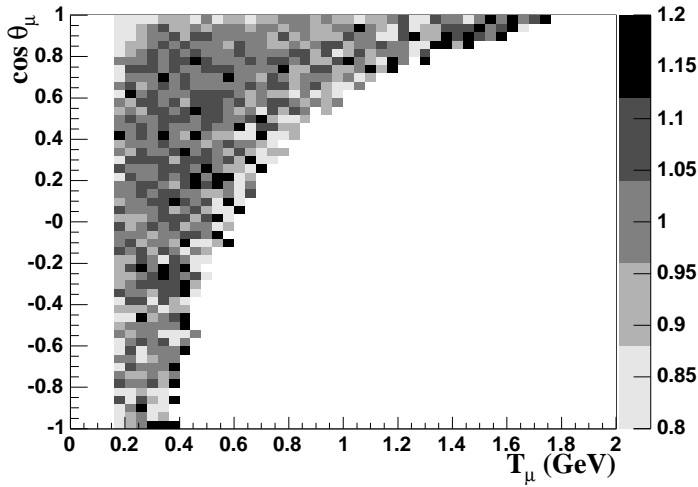}
\caption{Ratio of data/simulation as a function of muon kinetic energy and
angle after CCQE model adjustments. The simulation has been 
normalized to the data. The $\chi^2/\mathrm{dof}=45.1/53$. Compare to 
Figure~\ref{fig:prl_fig1}.}
\label{fig:prl_fig3}
\end{figure}

Figure~\ref{fig:prl_fig3} shows the agreement between data and 
simulation after incorporation of the $M_A$ and $\kappa$ values from 
the $Q^2$ fit to MiniBooNE data. Comparing to Figure~\ref{fig:prl_fig1}, 
the improvement is substantial and the data are well-described throughout 
the kinematic phase space. 

Table~\ref{table:prl_tab1} shows the contributions to the systematic 
uncertainties on $M_A$ and $\kappa$. The detector model uncertainties
dominate the error in $M_A$ due to their impact on the energy and
angular reconstruction of CCQE events in the MiniBooNE detector. The
dominant error on $\kappa$ is the uncertainty in the $Q^2$ shape
of background events.
This error (not included in the contour of Figure~\ref{fig:prl_fig2})
is evaluated in a separate fit, where MiniBooNE CC $1\pi^+$ data are used
to set the background instead of the event generator prediction, and then
added in quadrature.

\begin{table}
\caption{Uncertainties in $M_A^{\rm eff}$ and $\kappa$ from the fit
to MiniBooNE $\numu$ CCQE data. %The total error is not a simple quadrature
%sum because of the correlation between the two parameters.}
The total error is not a simple quadrature sum due to correlations between
the $Q^2$ bins created by the systematic uncertainties.}
\begin{ruledtabular}
\begin{tabular}{ccc}
error source&$\delta M_A^{\rm eff}$&$\delta \kappa$ \\
\hline
data statistics              &0.03             &0.003         \\ 
neutrino flux                &0.04             &0.003         \\
neutrino cross sections      &0.06             &0.004         \\
detector model               &0.10             &0.003         \\ 
CC $\pi^+$ background shape   &0.02             &0.007         \\ \hline
total error                  &0.20             &0.011        \\ 
\end{tabular}
\end{ruledtabular}
\label{table:prl_tab1}
\end{table}

%------------- interpretation

The result reported here, $M_A^{\rm eff}=1.23\pm0.20$~GeV, is consistent 
with a recent K2K measurement on a water target, 
$M_A=1.20\pm0.12$~GeV~\cite{k2k-ma}. Both values are consistent with
but higher than the historical
value, $M_A=1.026\pm0.021$~GeV, set largely by deuterium-based bubble chamber 
experiments~\cite{past-ma}.  The $M_A$ value reported here should
be considered an ``effective parameter'' in the sense that it may be 
incorporating nuclear effects not otherwise included in the RFG model. 
In particular, it may be that a more proper treatment of the nucleon 
momentum distribution in the RFG would yield an $M_A$ value in closer agreement
to that measured on deuterium. Future efforts will therefore explore how 
the value of $M_A$ extracted from the MiniBooNE data is altered  upon 
replacement of the RFG model with more advanced nuclear 
models~\cite{new-model}.

%-------- conclusions -----------------------------------------------------

In summary, modern quasi-elastic scattering data on nuclear targets are 
revealing the inadequacies of present neutrino cross section simulations. 
Taking advantage of the high-statistics MiniBooNE $\numu$ CCQE  data, we 
have extracted values of an effective axial mass parameter, 
$M_A^{\rm eff}=1.23 \pm 0.20$~GeV, and a Pauli-blocking parameter, 
$\kappa=1.019\pm 0.011$, achieving substantially improved agreement with the 
observed kinematic distributions in this data set. Incorporation of both fit 
parameters allows, for the first time, a description of neutrino CCQE 
scattering on a nuclear target down to $Q^2 = 0$~GeV$^2$.

\bigskip
%-----------------------------------------------------------------------------

\begin{acknowledgments}
We wish to acknowledge the support of Fermilab, the Department of Energy,
and the National Science Foundation in the construction, operation, and 
data analysis of the MiniBooNE experiment. 
\end{acknowledgments}

%-----------------------------------------------------------------------------

%-----------------------------------------------------------------------------

\begin{references}
\bibitem{MB-PRL} A.A. Aguilar-Arevalo {\em et al.}, Phys. Rev. Lett. {\bf 98},
                 231801 (2007).
\bibitem{k2k-osc} S. Yamamoto {\em et al.}, Phys. Rev. Lett. {\bf 96},
                  181801 (2006).
% numu disappearance paper: M.H. Ahn {\em et al.}, Phys. Rev. {\bf D74}, 
% 072003 (2006).
\bibitem{smith-moniz} R.A. Smith and E.J. Moniz, Nucl. Phys. {\bf B43},
                      605 (1972); {\it erratum: ibid.} {\bf B101}, 547 (1975).
\bibitem{eb-pf} E.J. Moniz {\em et al.}, Phys.\ Rev.\ Lett.\  {\bf 26}, 
                445 (1971); the binding energy is 25~MeV from electron 
                scattering data plus 9~MeV from the $T=0,1$ splitting 
                plus Coulomb corrections.
\bibitem{bba} R. Bradford {\em et al.}, Nucl. Proc. Suppl. {\bf 159}, 127 
                 (2006).
\bibitem{past-qe} S.K. Singh and E. Oset, Nucl. Phys. {\bf A542}, 587 (1992).
\bibitem{GEANT4} S. Agostinelli {\em et al.}, Nucl. Instrum. Meth. {\bf A506},
                 250 (2003).
\bibitem{flux-param} J.R. Sanford and C.L. Wang, BNL AGS internal reports
                     11299 and 11479 (1967), unpublished.
\bibitem{harp} M.G. Catanesi {\em et al.}, Eur. Phys. J. {\bf C52},
               29 (2007). % hep-ex/0702024
\bibitem{e910} I. Chemakin {\em et al.},  arXiv:0707.2375v1 [nucl-ex],
               to be published in Phys. Rev. C.
% no, Jocelyn points out that this data was not used (only E910 and HARP)
%                  T. Abbott {\em et al.}, Phys. Rev. {\bf D45}, 3906 (1992);
%                  J.V. Allaby {\em et al.}, CERN 70-12 (1970);
%                  D. Dekkers {\em et al.}, Phys. Rev. {\bf B962}, 137 (1965); 
%                  G.J. Marmer {\em et al.}, Phys Rev. {\bf 179}, 1294 (1969);
%                  T. Eighten {\em et al.}, Nucl. Phys. {\bf B44} 333 (1972);
%                  Y.D. Aleshin {\em et al.}, ITEP-77-80 (1977);
%                  I.A. Vorontsov {\em et al.}, ITEP-88-11 (1988).
\bibitem{nuance} D. Casper, Nucl. Phys. Proc. Suppl. {\bf 112}, 161 (2002).
\bibitem{ashery} D. Ashery {\em et al.}, Phys. Rev. {\bf C23}, 2173 (1981);
                 H. Ejiri, Phys. Rev. {\bf C48}, 1442 (1993); 
                 F. Ajzenberg-Selove, Nucl. Phys. {\bf A506}, 1 (1990).
\bibitem{GEANT3} CERN Program Library Long Writeup W5013 (1993).
\bibitem{gcalor} C. Zeitnitz and T.A. Gabriel, Nucl. Instrum. Meth.
                 {\bf A349}, 106 (1994).
\bibitem{ieee} B.C. Brown {\em et al.}, IEEE Nucl. Sci. Symp.
               Conf. Rec. {\bf 1}, 652 (2004).
\bibitem{muoncap} T. Suzuki {\em et al.}, Phys. Rev. {\bf C35}, 2212 (1987).
\bibitem{pseudo} K.F. Liu {\em et al.}, Phys. Rev. Lett. {\bf 74}, 2172 (1995).
\bibitem{ga} W.M. Yao {\em et al.}, J. Phys. {\bf G33}, 1 (2006).
\bibitem{past-ma} V. Bernard {\em et al.}, J. Phys. {\bf G28}, R1 (2002).
\bibitem{k2k-ma} R. Gran {\em et al.}, Phys. Rev. {\bf D74}, 052002 (2006).
\bibitem{new-model} J.E. Amaro {\em et al.}, Phys. Rev. {\bf C71}, 015501 
             (2005);  % 1st Donnelly paper on using e- data for nu predictions
             T. Leitner {\em et al.}, Phys. Rev. {\bf C73}, 065502 
             (2006);  % Alvarez paper on CC neutrino-nucleus scattering
%             O. Benhar and D. Meloni, hep-ph/0610403; 
             O. Benhar {\em et al.}, Phys. Rev. {\bf D72}, 053005 (2005);
             S. Ahmad {\em et al.}, Phys. Rev. {\bf D74}, 073008 (2006). 
             % Ahmad reference is the Singh 1pi paper which includes
             % calculation of modifications to the Smith-Moniz QE predictions 
             % due to RPA correlations in the QE xsec
\end{references}
\end{document}